\documentclass{article}

\usepackage{epsf}

\begin{document}
\begin{center}
\begin{LARGE}
Numerical Simulations of Galaxy Formation:\\
       Cooling, Heating, Star Formation
\end{LARGE}

\bigskip

\begin{large}
{\bf Anatoly Klypin$^1$, Stefan Gottl\"ober$^{1,2}$, \& Gustavo Yepes$^3$}
\end{large}
\bigskip

{$^1$Department of Astronomy, NMSU, Las Cruces, NM 88001}

{$^2$AIP, An der Sternwarte 16, D-14482 Potsdam, Germany}

{$^3$Departamento de Fisica Teorica C-XI, UNAM, Cantoblanco 28049, 
Madrid, Spain}
\end{center}

\noindent{\it 
Talk presented at the 12 th Postdam Cosmology
    Workshop, ``Large Scale Structure: Tracks and Traces'', World Scientific 1998}

\section{Abstract}

Formation of luminous matter in the Universe is a complicated process,
which includes many processes and components.  It is the vastly
different scales involved in the process (from star formation on few
parsec scales to galaxy clusters and superclusters on megaparsecs
scales) and numerous ill-understood processes, which make the whole
field a maze of unsolved, but exciting problems.  We present new
approximations for numerical treatment of multiphase ISM forming
stars. The approximations were tested and calibrated using
$N$-body+fluid numerical simulations. We specifically target issues
related with effects of unresolved lumpinesses of the gas.  We show
that the degree of freedom is much smaller than  naively 
expected because of self-regulating nature of the process of global star
formation.  One of the problems of numerical simulations is related
with the uncertainties of approximation of the supernovae (SN)
feedback. It is often assumed that the feedback is mainly due to
momentum transferred by SN in to the ISM.  We argue that this may not be
true.  We present a realistic example of gas actively forming stars
with short cooling time for which the SF feedback is important, but the
kinetic energy of the gas motion due to SN is negligibly small as
compared with the thermal energy of gas.

\section{Introduction}

Numerical $N$-body+hydro simulations are important
tools for making theoretical predictions for galaxy formation in the
expanding Universe.  Unfortunately, the simulations are mired  by two
basic problems. i) Results on scales of interest (kpc and larger)
crucially depend on much smaller (pc) scales. This is very different from
$N$-body problem where the situation on large scales does not depend on
what happens on small scales. The reason why small scales are important
lies in the nature of the cooling and heating of the gas. Small-scale
lumpiness changes the gas cooling rate. The gas heating is defined by
how the star formation proceeds on very small scales, not simply by
the average star formation rate. ii) There is no theory of star
formation (SF). Observations regarding SF are very useful, but they do
not cover all important situations. 

Nevertheless, the situation is not that grim as it looks. Not all
details of the small-scale processes are really important. Recently,
Yepes et al.(1997)\cite{Yepes} presented  realistic simulations of the galaxy
formation in which  different parameters (e.g., cooling rates,
resolution, SN feedback) were changed dramatically with
the goal to find out how the final results are sensitive on the
parameters.  The final global results (for example, total luminosities
and SFR) were remarkably stable when the cooling rates changed by a
factor of ten and when the resolution changed by a factor of two.
The results for small galaxies were very sensitive to the feedback
parameter, while large galaxies were insensitive even to the SN
feedback. Some of the dependencies are easy to understand, but some are
more difficult. For example, if the cooling time is very short (as
typically is the case), the cooling time is not important. Another
reason for stability lies in the nonlinear response of the ISM to the
SF. The ISM is typically found in a self-regulating regime when the
rate of infall of a fresh gas from outside is defined by the heating of
the gas by newly formated gas. If the infall is large, the SF increases
(because the gas cools and becomes available for SF), this results in
larger heating of the gas by young stars and SN, which stops the
infall. This  self-regulating evolution of the ISM results in effective
``canceling'' of the number of free parameters -- results are sensitive
only to a small number of combinations of the parameters.

The following example illustrates the idea. Let $x$ and $y$ be two
variables, which describe the situation (e.g., density and temperature
of gas). They depend one on the other in a simple, but nonlinear way,
which involves four parameters $a_1,a_2,$ $b_1, b_2$. The evolution of
the system is defined by two differential equations. The system has the
same basic structure as a real system: $\frac{dx}{dt} =a_1x-b_1y^2$,
$\frac{dy}{dt} =a_2y-b_2x^2$.  The system depends on four parameters
$a_i, b_i$ and on two initial conditions: $x_0$, $y_0$. Thus, six
``free parameters'' define the evolution of the system in general. With
six free parameters one naively would expect that any final
configuration is possible. But in the regime of self-regulating
evolution, $\frac{dx}{dt}\approx
\frac{dy}{dt}\approx 0$, and the solution depends only on two
parameters $\alpha =(a_1/b_1)^{1/3}$ and $\beta =(a_2/b_2)^{1/3}$, and
it does not depend on initial values of $x$ and $y$.

\section{Cooling of gas with multiple phases }

The equations describing the evolution of a two-phase medium of hot  gas,
which emits radiation, and cold clouds ($T_c=10^4K$), which effectively
have stopped cooling, but
are capable of producing luminous matter (``stars''), can be written in the
form\cite{Yepes}: 
\begin{equation}
\rho_h\frac{d\epsilon_h}{dt}=
\frac{\beta}{t_*}\left[ \epsilon_{SN} -A(\epsilon_h -\epsilon_c)\right]
 -\alpha\rho_h^2\frac{\Lambda(T_h)}{\mu^2m_H^2}, \quad
\epsilon_{SN} =10^{51}erg/22M_{\odot},
\end{equation}
\begin{equation}
\frac{d\rho_h}{dt}=
-\frac{C-\alpha}{\epsilon_h
-\epsilon_c}\rho_h^2\frac{\Lambda(T_h)}{\mu^2m_H^2}
+\frac{A\beta\rho_c}{t_*}, \quad
\beta\approx 0.12,
\end{equation}
\begin{equation}
\frac{d\rho_{gas}}{dt}= -\frac{1-\beta}{t_*}\rho_c=-\frac{\rho_*}{dt}
\quad \rho_{gas}=\rho_h+\rho_c,
\end{equation}
\noindent where indices $h$, $c$, and $gas$ refer to the hot, cold, and
the total gas components; $kT=(\gamma-1)m_H\mu_m\epsilon$. Free
parameters $\alpha$ and $C$ were calibrated using numerical
simulations: $\alpha=0.95$, $C=2$. The feedback parameter must be in
the range $A=50-200$.
In order to test our approximations for cooling of multiphase gas we
run few numerical simulations of evolution of initially slightly
inhomogeneous gas without gravity and without SF. Simulation box is
small: 500pc-3kpc. For a real cosmological simulation the whole box
would be one resolution element. We use $64^3$ PPM code to run the
simulations. Initial conditions were either small random gaussian
fluctuations with RMS=0.15 or a simple sine-wave perturbation
$\delta\rho\propto \sin(x)\sin(y)\sin(z)$. Two
combinations of the size of the box and the gas density were chosen: i)
gas cools very fast (no motion of gas) or ii) the gas cools slowly, and
it has time to produce motion of gas across the box. Results are
presented in Figure 1.

\begin{figure}
\epsfxsize=0.9\hsize
\caption{Cooling of gas in the regime of thermal instability. Full
curves are for gas with initial random distribution of density
fluctuations with the rms $\sigma=0.15$. Initially the gas is  in the
pressure balance and $T=10^6K$. The dot-dashed curves are for gas with
one sine-wave perturbation of density with the same rms fluctuation as
for the full curves. As the gas cools the thermal instability
develops. It results in the formation of very dense cold lumps and hotter
gas with very low density. The long-dashed curves show evolution of a
homogeneous gas with the same initial average density and
temperature. The dotted curves show results for our multiphase
treatment. It clearly provides much better approximation as compared
with the homogeneous gas.}
\centering\leavevmode
\epsfbox{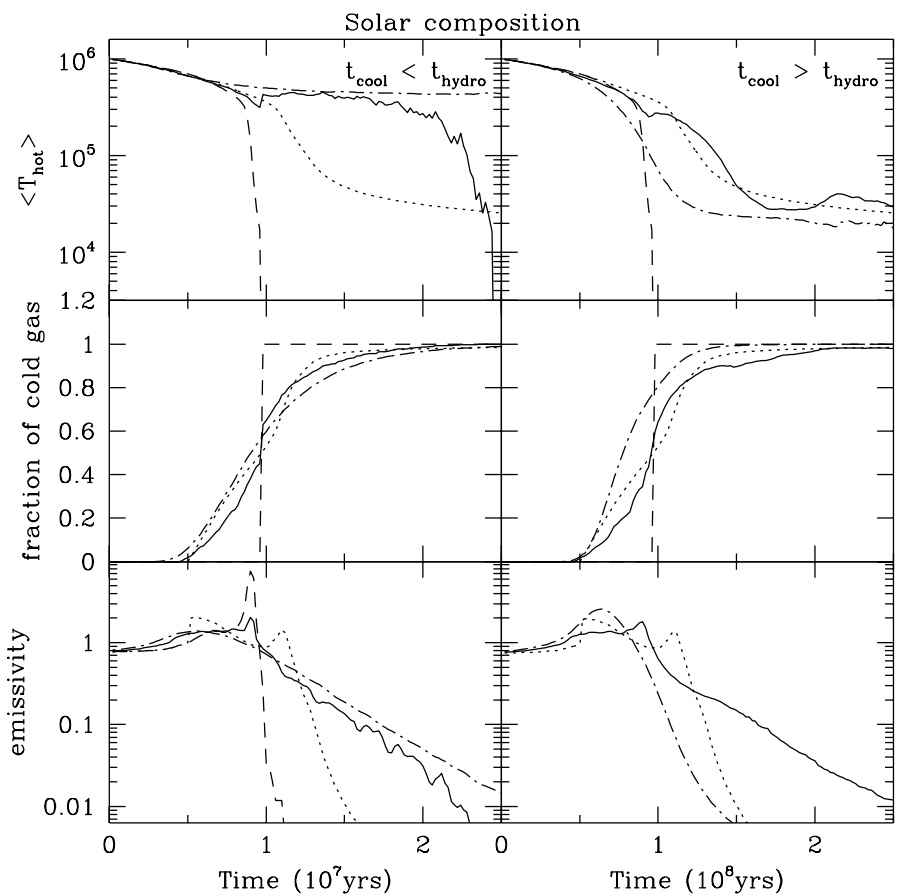}
\end{figure}

We also used the initial conditions to make runs with the SF and
feedback included. In all cases we observed that after some initial
evolution the system settles in a regime of steady ``burning'' of gas
into luminous matter (``stars''). In that regime the energy produced by
stars is almost equal to the energy radiated by the gas with the
kinetic energy of the gas always being much smaller than the thermal
gas energy. In the regime of the self-regulating SF the gas temperature
appears to be a constant, which can be found analytically:
\begin{equation}
\epsilon_h=\frac{(C-\alpha)\beta\epsilon_{SN}}
                             {C(1-\beta-A\beta)} + \epsilon_c
                    \approx \frac{(C-\alpha)}{C}\frac{\epsilon_{SN}}{A}
	\approx\frac{\epsilon_{SN}}{A}
\end{equation}

Figure 2 presents an example of the evolution of the multiphase gas with the
supernovae feedback. The SF was delayed by one cooling time in order
to increase the fraction of energy in the form of gas motion. Even in
this extreme case, the kinetic energy was much smaller than the thermal
energy of the gas. After $2\times 10^7$yrs the system was already in
the regime of self-regulating SF. The total energy  released by
forming stars was almost equal to the energy radiated by the gas with
the kinetic energy being much smaller than each of the energies. The
temperature of the gas was close to the value predicted by equation
(4). Note that vastly different initial conditions (one sine-wave or
1/4 million independent fluctuations) resulted in very similar final
states of the gas and almost identical  rates of conversion of gas to stars. 

We thank Volker Muller and the organizing committee of the Potsdam
Cosmology Workshop for their hospitality. This work is supported by NSF
grant AST-9319970 and NASA grant NAG-5-3842.

\begin{figure}
\epsfxsize=1.0\hsize
\caption{Evolution of average gas properties in the case of active star
formation and the supernovae feedback. 3D PPM hydro code with 50pc
resolution was used. The star formation was allowed only after
$10^7$yrs. The effective time-scale for SF $t_{* eff}=1.5\times
10^8$yrs was much longer than the
minimum $t_*=5\times 10^7$yrs imposed by the code. }
\centering\leavevmode
\epsfbox{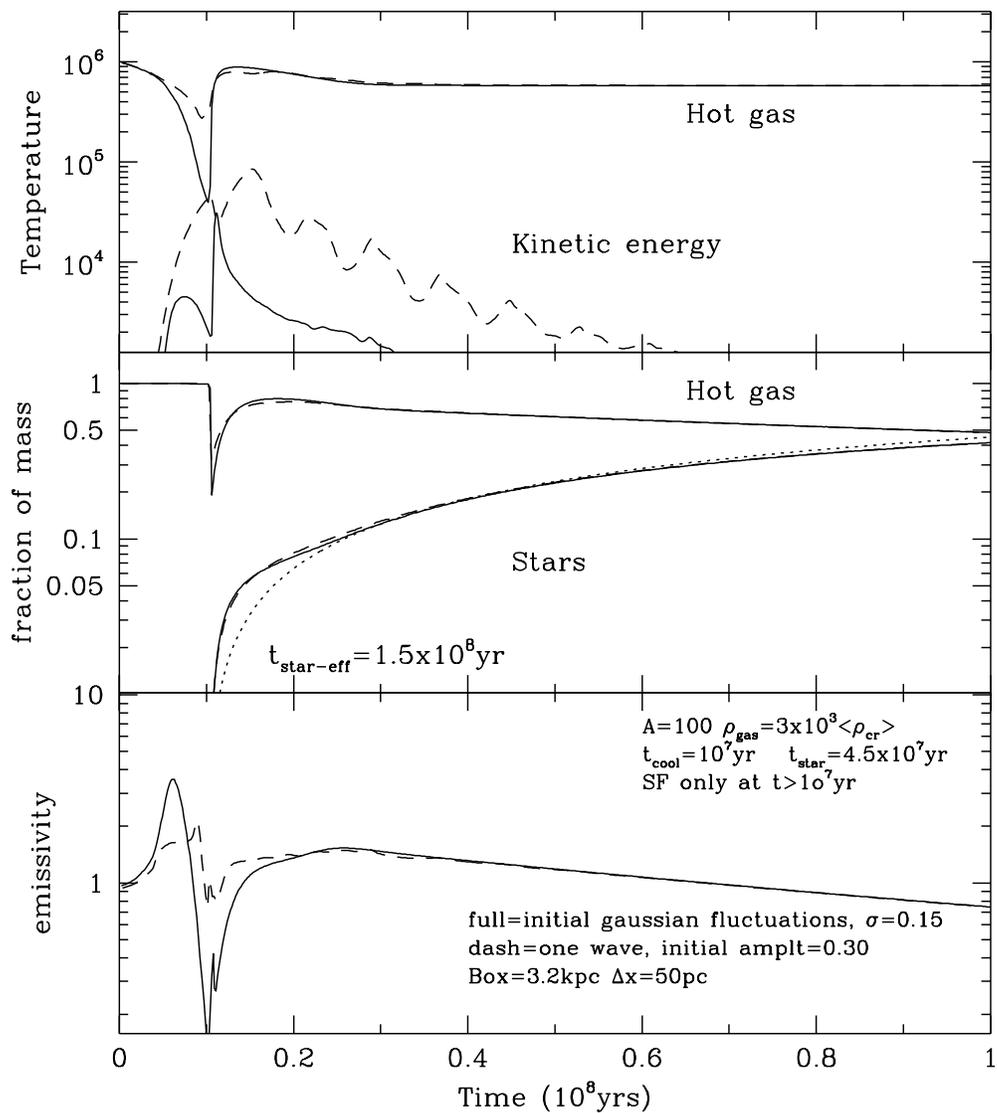}
\end{figure}
%\section*{References}


\begin{thebibliography}{99}
\bibitem{Yepes} Yepes, G., Kates, R., Khokhlov, A., Klypin, A., 1997,
MN, 284, 235.

\end{thebibliography}
\end{document}